# Coexistence of Superconductivity and Antiferromagnetism in Topological Magnet MnBi$_2$Te$_4$ Films

Wei Yuan[1,6], Zi-Jie Yan[1,6], Hemian Yi[1], Zihao Wang[1], Stephen Paolini[1], Yi-Fan Zhao[1], Ling-Jie Zhou[1], Annie G. Wang[1], Ke Wang[2], Thomas Prokscha[3], Zaher Salman[3], Andreas Suter[3], Purnima P. Balakrishnan[4], Alexander J. Grutter[4], Laurel E. Winter[5], John Singleton[5], Moses H. W. Chan[1], and Cui-Zu Chang[1]

[1]Department of Physics, The Pennsylvania State University, University Park, PA 16802, USA

[2]Materials Research Institute, The Pennsylvania State University, University Park, PA 16802, USA

[3]Laboratory for Muon Spectroscopy, Paul Scherrer Institute, 5232 Villigen PSI, Switzerland

[4]NIST Center for Neutron Research, National Institute of Standards and Technology, Gaithersburg, MD 20899, USA

[5]National High Magnetic Field Laboratory, Los Alamos, NM 87544, USA

[6]These authors contributed equally: Wei Yuan and Zi-Jie Yan

Corresponding author: cxc955@psu.edu (C.-Z. C.).

**Abstract:** The interface of two materials can harbor unexpected emergent phenomena. One example is interface-induced superconductivity. In this work, we employ molecular beam epitaxy to grow a series of heterostructures formed by stacking together two non-superconducting antiferromagnetic materials, an intrinsic antiferromagnetic topological insulator MnBi$_2$Te$_4$ and an antiferromagnetic iron chalcogenide FeTe. Our electrical transport measurements reveal interface-induced superconductivity in these heterostructures. By performing scanning tunneling microscopy and spectroscopy **measurements, we observe a proximity-induced superconducting gap on the top surface of the MnBi$_2$Te$_4$ layer, confirming the interaction between superconductivity and antiferromagnetism in the MnBi$_2$Te$_4$ layer. Our findings will advance the fundamental inquiries into the topological superconducting phase in hybrid devices and provide a promising platform for the exploration of chiral Majorana physics in MnBi$_2$Te$_4$-based heterostructures.**

**Main text:** The vitality and utility of condensed matter physics are the results of and driven by at first glance unexpected, emergent phenomena when constituent atoms, molecules, and material systems are brought together under suitable conditions. When two materials are stacked together, the resultant interface between them sometimes shows macroscopic quantum phenomena. This process of bringing out new phenomena and new electronic states of matter at the interface of two different materials is known as interface engineering. For example, the interfaces between two II-VI or III-V compound semiconductors host a two-dimensional (2D) electron gas that exhibits the celebrated integer and fractional quantum Hall effect[1,2]; when two insulating complex oxides, LaAlO$_3$ and SrTiO$_3$, are put together, superconductivity is found at their interface[3-5]. Until now, interfacial superconductors that have been searched for and identified originate predominantly from two nonmagnetic materials[6-8], intending to avoid the potentially deleterious pair-breaking effect caused by spin-flip scattering at the interface[9]. Remarkably, interface-induced superconductivity has recently been observed in heterostructures consisting of a ferromagnetic topological insulator (TI) Cr-doped (Bi,Sb)$_2$Te$_3$ and an antiferromagnetic iron chalcogenide FeTe (Ref.[10]). This surprising discovery motivates us to investigate whether interface-induced superconductivity can also be achieved by substituting the ferromagnetic TI with an antiferromagnetic TI.

MnBi$_2$Te$_4$, a tetradymite-type compound, has been demonstrated to be an intrinsic antiferromagnetic TI (Refs.[11-15]) (Figs. 1a and 1b). The layered antiferromagnetic order in MnBi$_2$Te$_4$ has a substantial influence on its topological properties. Because of the ferromagnetism within one septuple layer (SL), each helical surface state is expected to experience a magnetic exchange gap due to surface magnetization and thus contribute $e^2/2h$ to the total conductance. Depending on whether the MnBi$_2$Te$_4$ film has an even or odd number of SLs, the top and bottom surface magnetizations will be antiparallel or parallel and the Hall conductance from the two surfaces will cancel or sum. The total conductance is $e^2/h$ in odd SLs [i.e., the quantum anomalous Hall (QAH) state] and 0 in even SLs (i.e., the axion insulator state) under zero magnetic field[11,12,15-22]. Both the QAH and axion insulator states have recently been observed in manually exfoliated odd- and even-SL MnBi$_2$Te$_4$ devices[15,21,22], respectively. The thin films of MnBi$_2$Te$_4$ were also achieved by molecular beam epitaxy (MBE) growth[13,23-29]. FeTe, an antiferromagnetic iron chalcogenide that has been intensively studied in both bulk and film forms, is not superconducting without elemental doping[30,31] or tensile stress[32]. Since MnBi$_2$Te$_4$ and Cr-doped (Bi,Sb)$_2$Te$_3$ share a similar layered rhombohedral structure, a similar in-plane lattice constant, and similar growth conditions[15], these similarities and the success in growing Cr-doped (Bi,Sb)$_2$Te$_3$/FeTe heterostructures[10] prompt us to grow MnBi$_2$Te$_4$/FeTe heterostructures to search for superconductivity at the antiferromagnet/antiferromagnet interface (Figs. 1a and 1b).

In this work, we grow the heterostructures by stacking, in MBE growth, $m$ SL MnBi$_2$Te$_4$ films and $n$ unit-cell (UC) FeTe films. Below we denote $m$ SL MnBi$_2$Te$_4$/$n$ UC FeTe as the ($m$, $n$) heterostructure. The MnBi$_2$Te$_4$/FeTe interface is found to be atomically sharp. By performing electrical transport and scanning tunneling microscopy/spectroscopy (STM/S) measurements, we observe interface-induced superconductivity in MnBi$_2$Te$_4$/FeTe heterostructures and demonstrate the coexistence of superconductivity and antiferromagnetism in the $MnBi_2Te_4$ layer. Moreover, we find that the upper critical magnetic field of the emergent superconductivity is extremely high (>39.5 T) and nearly isotropic at $T = 1.5$ K, which may enable the coexistence of superconductivity and antiferromagnetism in our $MnBi_2Te_4$/FeTe heterostructures.

All $MnBi_2Te_4$/FeTe heterostructures are grown on heat-treated ~0.5 mm thick $SrTiO_3$(100) substrates in a commercial MBE chamber (Omicron Lab10) (Methods). The MBE growth is monitored using *in-situ* reflection high-energy electron diffraction (RHEED) (Supplementary Fig. 1). The STM/S measurements are performed in a Unisoku 1300 system (310 mK, 11 T) with a base vacuum better than $3 \times 10^{-10}$ mbar. The electrical transport studies are conducted in two Physical Property Measurement Systems (Quantum Design DynaCool, 1.7 K, 9 T/14 T) and a capacitor-driven 65 T pulsed magnet at the National High Magnetic Field Laboratory (NHMFL), Los Alamos. The mechanically scratched Hall bars are used for electrical transport measurements. More details about the MBE growth, sample characterization, STM/S, and electrical transport measurements can be found in Methods.

Both $MnBi_2Te_4$ and FeTe crystals are antiferromagnetic and inherently non-superconducting at low temperatures. $MnBi_2Te_4$ has a lattice structure composed of Te-Bi-Te-Mn-Te-Bi-Te SLs. The lattice structure of $MnBi_2Te_4$ can be viewed as intercalating a MnTe bilayer into a $Bi_2Te_3$ quintuple layer. Below its *Néel* temperature $T_N$ ~24 K, $MnBi_2Te_4$ displays a layered (i.e., A-type) antiferromagnetic order due to two Mn sublattices[13-15,33,34]. As noted above, the magnetic moments of Mn atoms couple ferromagnetically within each SL, but antiferromagnetically between adjacent SLs. The magnetic moment is aligned along the *c*-axis (Fig. 1a). For FeTe, each UC comprises one layer of Fe atoms sandwiched by two layers of Te atoms. The spins of the Fe atoms align diagonally across the Fe-Fe square lattice, giving rise to a bicollinear antiferromagnetic order (Fig.

1b)[35]. The formation of its antiferromagnetic order is accompanied by a structural phase transition, with the structure changing from a tetragonal to a monoclinic phase. Its *Néel* temperature $T_N$ is in a range of 60~70 K (Refs.[36-39]). Both MnBi$_2$Te$_4$ and FeTe are layered materials but have different lattice structures (i.e., rhombohedral *vs* tetragonal) (Figs. 1a and 1b). We still obtain high-quality MnBi$_2$Te$_4$/FeTe heterostructures in our experiments.

We first characterize the MnBi$_2$Te$_4$/FeTe heterostructures. Supplementary Fig. 1a displays the RHEED patterns of heat-treated SrTiO$_3$(100) substrate. The reconstructed RHEED patterns of the SrTiO$_3$(100) substrate show an atomically flat surface and hence suitable for MBE growth of the FeTe films. The sharp and streaky "1×1" RHEED patterns for MnBi$_2$Te$_4$ and FeTe films demonstrate the highly ordered crystal structures and the sharp interface between these two layers (Supplementary Figs. 1b and 1c). Figure 1c shows the X-ray diffraction (XRD) spectra of the (9, 35) heterostructure. Clear peaks corresponding to the MnBi$_2$Te$_4$ and FeTe layers, as well as SrTiO$_3$(100) substrate are observed, confirming the high crystalline quality of our MnBi$_2$Te$_4$/FeTe heterostructures. Our cross-sectional scanning transmission electron microscopy (STEM) measurements resolve the highly ordered SL structure of MnBi$_2$Te$_4$ and the trilayer structure of FeTe, showing an atomically sharp interface (Fig. 1d and Supplementary Fig. 2). We note that the atomic structures of MnBi$_2$Te$_4$ and FeTe layers cannot be simultaneously resolved, which indicates a misorientation due to their different lattice structures and symmetries. The STEM images reveal an in-plane rotation angle of ~15° between MnBi$_2$Te$_4$ and FeTe (Fig. 1d and Supplementary Fig. 2).

To investigate the interface-induced superconductivity in heterostructures formed by stacking together antiferromagnetic MnBi$_2$Te$_4$ and FeTe layers, we maintain the thickness of the FeTe layer at $n = 35$ and grow a series of the ($m$, 35) heterostructures with $0 \leq m \leq 10$. The specific value of $m$ is determined by the growth duration and then calibrated by the atomic force microscopy measurements. We perform electrical transport measurements on these ($m$, 35) heterostructures. For the $m = 0$ sample, i.e., the 35 UC FeTe layer, the $R_{xx}$-$T$ curve shows a non-superconducting behavior (Fig. 2a). This observation is also confirmed by the absence of the superconducting gap in the d$I$/d$V$ spectra measured on the top surface of the FeTe layer (Fig. 3c). A hump feature, which is usually associated with the paramagnetic-to-antiferromagnetic phase transition (i.e., $T_N$), is observed at $T \sim 62$ K (Fig. 2a). Once the MnBi$_2$Te$_4$ layer with a thickness of 0.2 SL is deposited onto the 35 UC FeTe layer, superconductivity emerges. For the $m = 0.2$ sample, the zero-resistance state appears below $T_{c,0} \sim 3.6$ K, while the superconducting onset temperature $T_{c,\text{onset}}$ is $\sim 8.6$ K. Here $T_{c,\text{onset}}$ is defined when $R_{xx}$ begins its drop from the normal state value. With a further increase in $m$, $T_{c,\text{onset}}$ exhibits a slight increase, while $T_{c,0}$ shows a much more significant enhancement. Specifically, for the $m = 0.4$ sample, $T_{c,0} \sim 8.6$ K and $T_{c,\text{onset}} \sim 10.6$ K. Both $T_{c,0}$ and $T_{c,\text{onset}}$ values reach saturation for $m \geq 1$ (Fig. 2a).

Moreover, we observe that the $T_N$ hump feature broadens and shifts to a higher temperature with increasing $m$. This suggests that the suppression of antiferromagnetism in the FeTe layer is linked to the emergence of interface-induced superconductivity. For $m \geq 1$, the $T_N$ hump feature shifts to an even higher temperature and eventually reaches saturation. We note that for the $m = 7$ sample, both $T_{c,\text{onset}}$ and $T_{c,0}$ reach maximum values of $\sim 11.1$ K and $\sim 9.2$ K, respectively (Fig. 2a). For a straightforward comparison of different samples, we define the superconducting transition temperature $T_c$ as the temperature at which $R_{xx}$ drops to $\sim 50\%$ of its normal state value of $T = 20$ K and plot the $T_c$ values as a function of $m$ (Fig. 2b). For $m \leq 1$, we observe a significant increase in $T_c$ with increasing $m$. However, for $m \geq 1$, the $T_c$ values show a weak fluctuation, consistent with the above analysis.

To demonstrate the presence of antiferromagnetism in the MnBi$_2$Te$_4$ layer, we carry out Hall measurements on the (7, 35) heterostructure at different temperatures (Figs. 2c and 2d, Supplementary Fig. 4). At $T = 8$ K, which is below its $T_{c,0}$ ~9.2 K, the Hall resistance $R_{yx}$ is vanishing, which is a result of the zero-resistance state of the superconducting phase (Fig. 2c). At $T = 12$ K, which is slightly greater than $T_{c,onset}$ ~11.1 K, a small non-square hysteresis loop appears near zero magnetic field (Fig. 2d), which is presumably induced by the coexistence of the uncompensated magnetization in the dominant MnBi$_2$Te$_4$ phase (Refs.[40,41]) and the minor ferromagnetic Mn-doped Bi$_2$Te$_3$ phase[25]. A kink feature, associated with the spin-flop transition, has been employed in some studies to identify the antiferromagnetism in the MBE-grown MnBi$_2$Te$_4$ layer [13,15,23-29]. This kink feature is not resolvable in the (7, 35) heterostructure (Fig. 2d), possibly due to the following two reasons: (*i*) the measurement temperature needs to exceed its $T_{c,onset}$ ~11.1 K; and (*ii*) the high conductivity of the bottom 35 UC FeTe layer may divert current flow within the heterostructure sample. However, at $T = 12$ K, the overall shape of the Hall traces for the (7, 35) heterostructure (Figs. 2d and Supplementary Fig. 4) closely resembles that of 7 SL MnBi$_2$Te$_4$ without the bottom FeTe layer (Supplementary Fig. 5f). Furthermore, the non-square hysteresis loop near zero magnetic field disappears at $T = 25$ K (Supplementary Fig. 4e), consistent with the $T_N$ value of 7 SL MnBi$_2$Te$_4$. Both observations indicate the persistence of antiferromagnetism within the MnBi$_2$Te$_4$ layer in the heterostructure samples.

Besides varying $m$ at $n = 35$, we also fix $m = 7$ and perform electrical transport measurements on the (7, $n$) MnBi$_2$Te$_4$/FeTe heterostructures with different $n$. For the $n = 0$ sample, i.e., 7 SL MnBi$_2$Te$_4$ film on SrTiO$_3$(111) substrate grown by the same recipe, a kink feature is observed at $T$ ~24 K, which suggests $T_N$ ~24 K of the 7SL MnBi$_2$Te$_4$ film (Supplementary Fig. 5a). For $T \leq 15$ K, $R_{xx}$ exhibits an abrupt increase, which is probably attributable to the insulating ground state induced by electron-electron interaction in TI thin films[42,43]. With a further increase in $n$, $R_{xx}$ gradually decreases, and the interface-induced superconductivity progressively becomes unmistakable (Supplementary Figs. 5b to 5e). As we noted above, the absence of the $T_N$ hump feature for $n \geqslant 7$ is probably due to the much higher conductivity of the bottom FeTe layer compared to the top 7 SL MnBi$_2$Te$_4$ layer.

The (7, 7) heterostructure shows an onset of superconductivity near ~10.5 K without ever reaching a zero-resistance state. This indicates a lack of true long-range superconducting coherence within this thinner FeTe film (Supplementary Fig. 5b). At $T$ = 12 K, the Hall traces of the (7, 7) heterostructure resemble those of the (7, 0) heterostructure but shrink by one order of magnitude (Supplementary Figs. 5f and 5g). The spin-flop transition kink feature is also observed near $\mu_0 H$ ~ ±4.0 T (Supplementary Fig. 5g). With a further increase in $n$, the conductivity of the FeTe layer is enhanced, and thus more electrical current flows through the bottom FeTe layer. Therefore, the Hall traces of the (7, $n$) heterostructures further shrink but the spin-flop transition kink feature is still observed at $T$ = 12 K (Supplementary Figs. 5f to 5g). These observations further confirm the antiferromagnetic property of the top 7 SL MnBi$_2$Te$_4$ layer.

To validate the proximity-induced superconducting gap on the top surface of the MnBi$_2$Te$_4$ layer, we perform low-temperature STM/S measurements on three ($m$, 35) heterostructures with $m$ = 0, 1, and 5 (Fig. 3a). Figure 3b shows the STM topographic image with atomic resolution of the (5, 35) heterostructure. We can clearly see hexagonal close-packed Te atoms on the top surface of the 5 SL MnBi$_2$Te$_4$ layer. The triangle-shaped dark features are the Mn/Bi anti-sites in the MnBi$_2$Te$_4$ layer[44]. On the top surface of the $m$ = 0 heterostructure, i.e., the 35 UC FeTe layer, no superconducting gap is observed near zero sample bias (Fig. 3c), consistent with its $R_{xx}$-$T$ curve (Fig. 2a).

Upon deposition of 1 SL MnBi$_2$Te$_4$ on the 35 UC FeTe layer, superconductivity appears. For the (1, 35) heterostructure, we observe a proximity-induced superconducting gap on the top surface of the 1 SL MnBi$_2$Te$_4$ layer (Fig. 3d, Supplementary Figs. 6 and 7). This observation is in good agreement with our transport results (Fig. 2). At $T = 310$ mK, the superconducting gap size is ~ 2.9 meV, which is determined from the fit using the Dynes formula (Supplementary Fig. 6)[45]. With increasing $T$, the proximity-induced superconducting gap gradually diminishes and ultimately vanishes at $T = 12.3$ K (Fig. 3d, Supplementary Figs. 7a and 7c). With a further increase in $m$, the proximity-induced superconductivity is attenuated on the top surface of the MnBi$_2$Te$_4$ layer. For the (5, 35) heterostructure, we observe a narrower and shallower superconducting gap at $T = 310$ mK (Fig. 3e). This observation further confirms that the superconducting gap observed on the top surface of the MnBi$_2$Te$_4$ layer is indeed induced by the proximity effect. With increasing $T$, this superconducting gap shrinks faster and disappears at $T = 3.0$ K (Fig. 3e, Supplementary Figs. 7b and 7c). Furthermore, our low energy muon spin relaxation (LE-μSR) measurements on the (7, 35) heterostructure also support a uniform antiferromagnetic order in the MnBi$_2$Te$_4$ layer, which is achieved above $T_{c,onset}$ and persists below $T_{c,0}$ (Supplementary Fig. 10). By combining electrical transport with STM/S and LE-μSR results, we demonstrate the coexistence of intrinsic antiferromagnetism and proximity-induced superconductivity in the MnBi$_2$Te$_4$ layer.

Next, we perform electrical transport measurements on the (9, 35) heterostructure under high magnetic fields to further our understanding of the coexistence of superconductivity and antiferromagnetism in MnBi$_2$Te$_4$/FeTe heterostructures. Figures 4a and 4b show the $R_{xx}$-$\mu_0 H$ curves at different temperatures with $\theta = 0°$ and $90°$, respectively. Here $\theta$ is defined as the angle between the normal direction of the sample plane and the direction of $\mu_0 H$ (Fig. 4c inset). We observe similar behaviors in these $R_{xx}$-$\mu_0 H$ curves at $\theta = 0°$ and $90°$ under different temperatures.

At $T = 1.5$ K, $R_{xx}$ remains zero until $\mu_0H$ reaches ~32.8 T. Subsequently, $R_{xx}$ dramatically increases and returns to the normal state near $\mu_0H$ ~45.0 T (Figs. 4a and 4b). To examine the anisotropy of interface-induced superconductivity in MnBi$_2$Te$_4$/FeTe heterostructures, we measure the $R_{xx}$-$\mu_0H$ curves of the (9, 35) heterostructure at different $\theta$ and $T = 1.5$ K. We find that the $R_{xx}$-$\mu_0H$ curves show nearly identical behavior under all $\theta$, implying the interface-induced superconductivity in MnBi$_2$Te$_4$/FeTe heterostructures is isotropic (Fig. 4c). We define the upper critical magnetic field $\mu_0H_{c2}$ as the magnetic field at which $R_{xx}$ drops to ~50% of its normal state value. The value of $\mu_0H_{c2}$ is ~39.5 T under different $\theta$ (Fig. 4d). The isotropic angle dependence suggests the interface-induced superconductivity is bulk-like in our MnBi$_2$Te$_4$/FeTe heterostructures. We note that these features are consistent with our recent findings in Cr-doped (Bi, Sb)$_2$Te$_3$/FeTe heterostructures[10], verifying that the emergent superconductivity originates from the FeTe layer. Moreover, we note that the large $\mu_0H_{c2}$ may be the key condition that allows for the coexistence of superconductivity and antiferromagnetism in the MnBi$_2$Te$_4$ layer. Further experimental and theoretical studies are needed to clarify this point.

To summarize, we employ MBE to synthesize MnBi$_2$Te$_4$/FeTe heterostructures and discover an emergent interface-induced superconductivity formed by stacking together two antiferromagnetic layers. By performing electrical transport and STM/S measurements, we demonstrate the coexistence of superconductivity and antiferromagnetism within the MnBi$_2$Te$_4$ layer. Furthermore, we find that the upper critical magnetic field of the interface-induced superconductivity in MnBi$_2$Te$_4$/FeTe heterostructures is large and isotropic, which may be responsible for the coexistence of superconductivity and antiferromagnetism. The MBE-grown MnBi$_2$Te$_4$/FeTe heterostructures with robust interface-induced superconductivity and atomically sharp interfaces provide a promising platform for the exploration of chiral Majorana physics [46-48] and thus constitute an important step toward scalable topological quantum computation.

**Methods**

**MBE growth**

All MnBi$_2$Te$_4$/FeTe heterostructures used in this work are grown in a commercial MBE chamber (Omicron Lab10) with a vacuum better than $2 \times 10^{-10}$ mbar. The insulating SrTiO$_3$(100) substrates are first soaked in ~80 °C deionized water for ~2 hours and then put in a diluted hydrochloric acid solution (~4.5% w/w) for ~2 hours. These SrTiO$_3$(100) substrates are annealed at ~974 °C for 3 hours in a tube furnace with flowing high-purity oxygen gas. Through these treatments, the surface of the SrTiO$_3$(100) substrate becomes passivated and shows a well-ordered reconstruction (Supplementary Fig. 1). We next load the heat-treated SrTiO$_3$ (100) substrates into our MBE chamber and outgas them at ~600 °C for ~1 hour. High-purity Mn (99.9998%), Bi (99.9999%), Fe (99.995%), and Te (99.9999%) are evaporated from Knudsen effusion cells. The growth temperatures are ~340 °C and ~270 °C for the FeTe and MnBi$_2$Te$_4$ layers, respectively. The growth rate is ~0.3 UC per minute for the FeTe layer and ~0.2 SL per minute for the MnBi$_2$Te$_4$ layer. Both are calibrated by measuring the thickness of the ($m$, $n$) heterostructures using both STEM and atomic force microscopy measurements. No capping layer is involved in our *ex-situ* electrical transport.

**XRD measurements**

The XRD measurements are performed using a Malvern Panalytical X'Pert3 MRD at room temperature. The copper $K_\alpha$ line at a wavelength λ ~1.5418 Å is used.

**STEM measurements**

The STEM measurements are performed on an FEI Titan³ G2 STEM operating at an accelerating voltage of 300 kV, with a probe convergence angle of 25 mrad, a probe current of 100 pA, and ADF detector angles of 42~244 mrad. See more STEM images of MnBi$_2$Te$_4$/FeTe heterostructures in Supplementary Fig. 2.

**ARPES measurements**

The ARPES measurements are performed in a chamber with a base vacuum of ~5 × 10$^{-11}$ mbar. The MnBi$_2$Te$_4$/FeTe heterostructure is transferred from the MBE chamber to the ARPES chamber without breaking the ultrahigh vacuum. A hemispherical Scienta R3000 analyzer is used. The photoelectrons are excited by a helium-discharged lamp with a photon energy of ~21.2 eV. The energy and angle resolutions are ~10 meV and ~0.1°, respectively. All ARPES measurements are performed at room temperature.

**Electrical transport measurements**

All MnBi$_2$Te$_4$/FeTe heterostructures for electrical transport measurements are scratched into a Hall bar geometry using a computer-controlled probe station. The effective area of the Hall bar device is ~1 mm × 0.5 mm. The electrical contacts are made by pressing indium spheres on the films. The electrical transport measurements under low magnetic fields (≤ 14 T) are conducted using two Physical Property Measurement Systems (Quantum Design DynaCool, 1.7 K, 9 T/14 T). Transport measurements under high magnetic fields (> 14 T) are conducted in a capacitor-driven 65 T pulsed magnet at the National High Magnetic Field Laboratory (NHMFL), Los Alamos. The excitation current is ~1 μA for $R_{xx}$-$T$ measurements and ~100 μA for Hall measurements, respectively.

**STM/S measurements**

The STM/S measurements are performed in a Unisoku 1300 system with a base vacuum better than 3 × 10$^{-10}$ mbar. The system incorporates a single-shot $^3$He cryostat to achieve a base temperature of ~310 mK. The maximum magnetic field of the system is ~11 T. Polycrystalline PtIr tips are used in our STM/S measurements. Before conducting STM/S measurements on MnBi$_2$Te$_4$/FeTe heterostructures, the PtIr tip is regularly conditioned on the MBE-grown Ag films. The *dI/dV* spectra are obtained by using the standard lock-in amplifier method by applying an additional small a.c. voltage at a frequency of ~983 Hz. All STM images are processed with WSxM 5.0 software [49].

**LE-μSR measurements**

The LE-μSR measurements are performed using the Low-Energy Muon Facility (LEM) at the Swiss Muon Source, Paul Scherrer Institute, Switzerland[50,51]. The sample used in our experiment consists of two identical 10×10 mm$^2$ pieces of the (7, 35) heterostructure, which are mounted with silver paint on a nickel-coated aluminum plate to minimize background effects. The samples are mounted in a helium flow cryostat (CryoVac, Konti), which is capable of maintaining the sample stage temperature within ±0.1 K of the target temperature. Fully polarized muons (μ$^+$) are accelerated to variable energies for implantation using high voltage. Upon implantation into the sample, the muon polarization is oriented within the film plane, and a transverse magnetic field of ~7.5 mT is applied perpendicular to the surface of the sample. Measurements are performed upon cooling with multiple energies at each temperature, specifically 1 keV, 1.5 keV, 2 keV, 3 keV, 4 keV, and 5 keV. The beam transport settings are ~12.0 kV. Before subsequent cooling, the sample is heated to ~200 K and the magnet is degaussed.

**Acknowledgments:** This work is primarily supported by the DOE grant (DE-SC0023113), including the MBE growth and PPMS (9T) measurements. The XRD, STEM and PPMS (14T) measurements are partially supported by the NSF-CAREER award (DMR-1847811) and the Penn State MRSEC for Nanoscale Science (DMR-2011839). The STM/S measurements are partially supported by the ARO grant (W911NF2210159). C. -Z. C. acknowledges the support from the Gordon and Betty Moore Foundation's EPiQS Initiative (Grant GBMF9063 to C. -Z. C.). Work done at NHMFL is supported by NSF (DMR-2128556) (J. S. and L. E. W.) and the State of Florida.

**Author contributions:** C. -Z. C. conceived and designed the experiment. W. Y., Z. -J. Y., H. Y., L. -J. Z., and A. G. W. performed the MBE growth and PPMS transport measurements. Z. -J. Y. performed the XRD measurements. K. W. and H. Y. carried out the STEM measurements. H. Y. and Z. -J. Y. performed the ARPES measurements. Z. W., S. P., and Y. -F. Z. performed STM/S measurements. H. Y., J. S., and L. E. W. performed the electrical transport measurements at NHMFL, Los Alamos. T. P., Z. S., A. S., P. P. B., and A. J. G. performed the μSR measurements. W. Y., Z. -J. Y., and C. -Z. C. analyzed the data and wrote the manuscript with input from all authors.

**Competing interests**: The authors declare no competing interests.

**Data availability:** The datasets generated during and/or analyzed during this study are available from the corresponding author upon request.

**Figures and figure captions:**

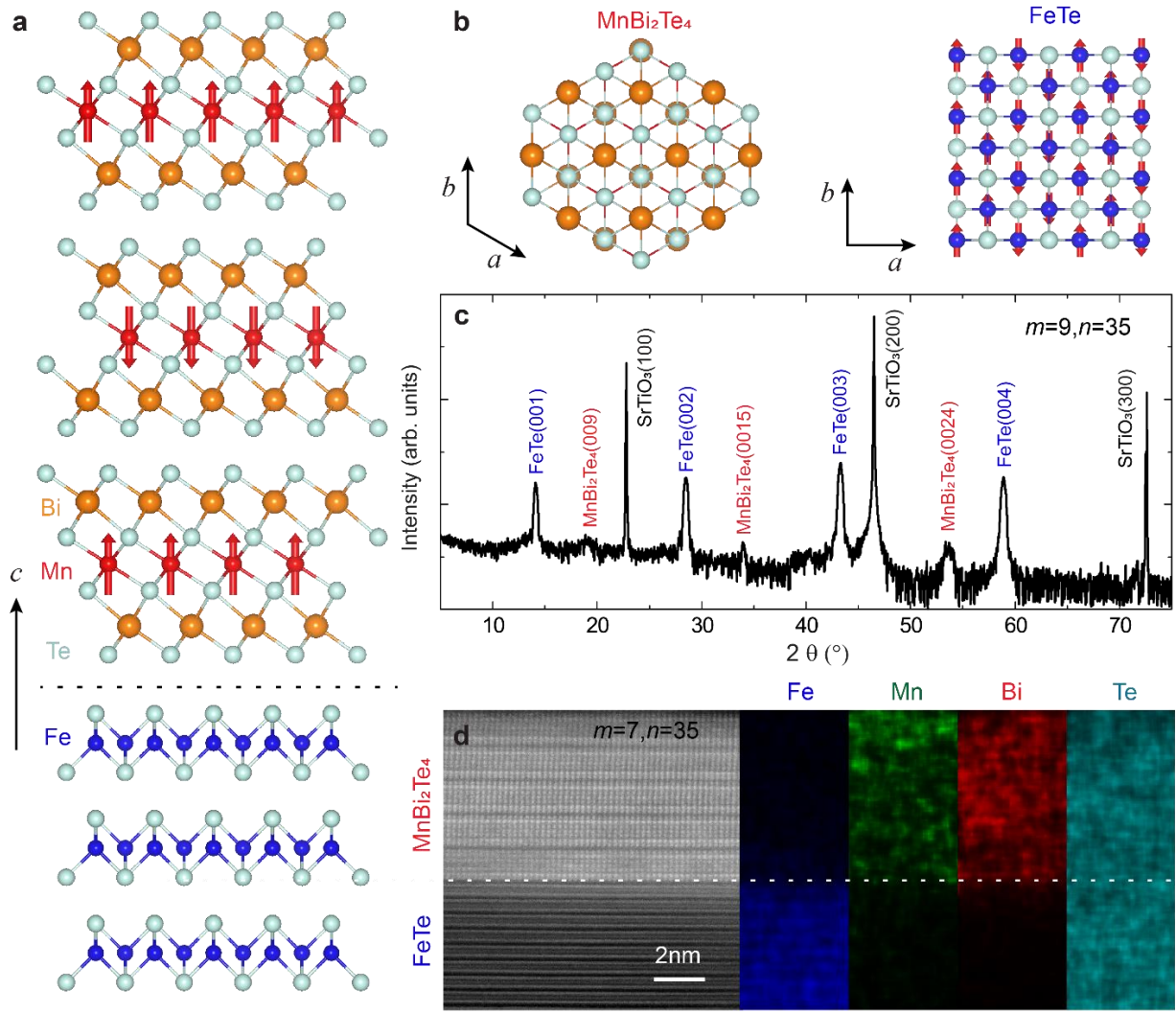

**Fig. 1| MBE-grown MnBi₂Te₄/FeTe heterostructures. a,** Side view of the MnBi$_2$Te$_4$/FeTe lattice structure. **b,** Top views of the MnBi$_2$Te$_4$ (left) and FeTe (right) lattice structures. The red arrows in (**a**) and (**b**) show the AFM orders in MnBi$_2$Te$_4$ and FeTe. **c,** XRD spectra of the (9, 35) heterostructure on heat-treated SrTiO$_3$(100). **d,** Cross-sectional STEM image and corresponding EDS maps of the (7, 35) heterostructure. A sharp interface is resolved between MnBi$_2$Te$_4$ and FeTe layers.

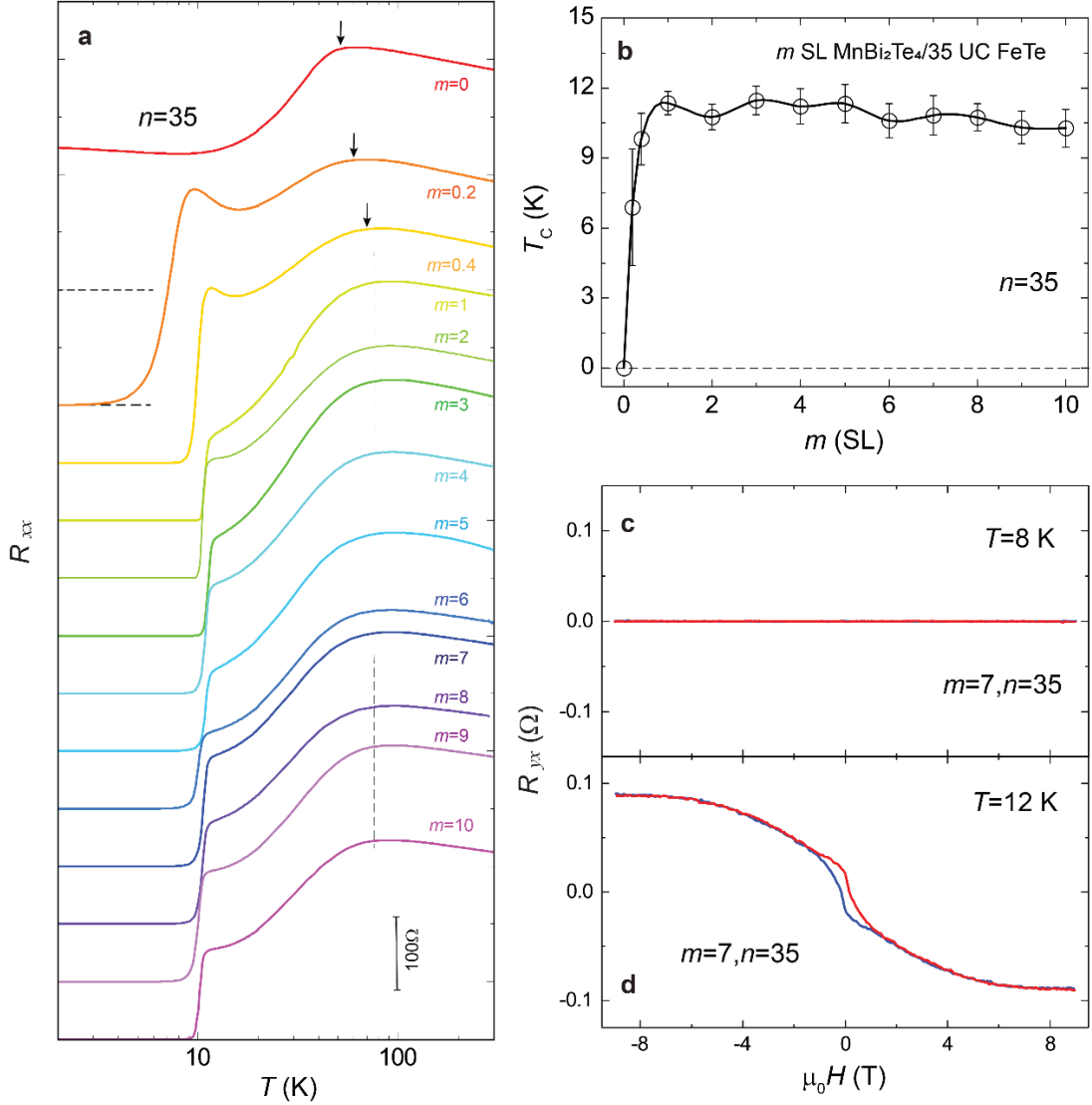

**Fig. 2| Interface-induced superconductivity in MnBi$_2$Te$_4$/FeTe heterostructures. a**, Temperature dependence of the sheet longitudinal resistance $R_{xx}$ of the ($m$, 35) heterostructures with $0 \leq m \leq 10$. The two horizontal dashed lines represent zero resistance in the $m = 0$ and $m = 0.2$ heterostructures. The three arrows and the vertical dashed lines indicate the hump features resulting from the paramagnetic-to-antiferromagnetic phase transition. **b,** $m$ dependence of the superconducting temperature $T_c$. The value of $T_c$ is defined as the temperature at which $R_{xx}$ drops to 50% of its normal state resistance. The error bar of each sample is estimated from the value difference between $T_{c,\,onset}$ and $T_{c,0}$. **c, d,** $\mu_0 H$ dependence of the Hall resistance $R_{yx}$ of the (7, 35) heterostructure at $T$=8 K (**c**) and $T$=12 K (**d**).

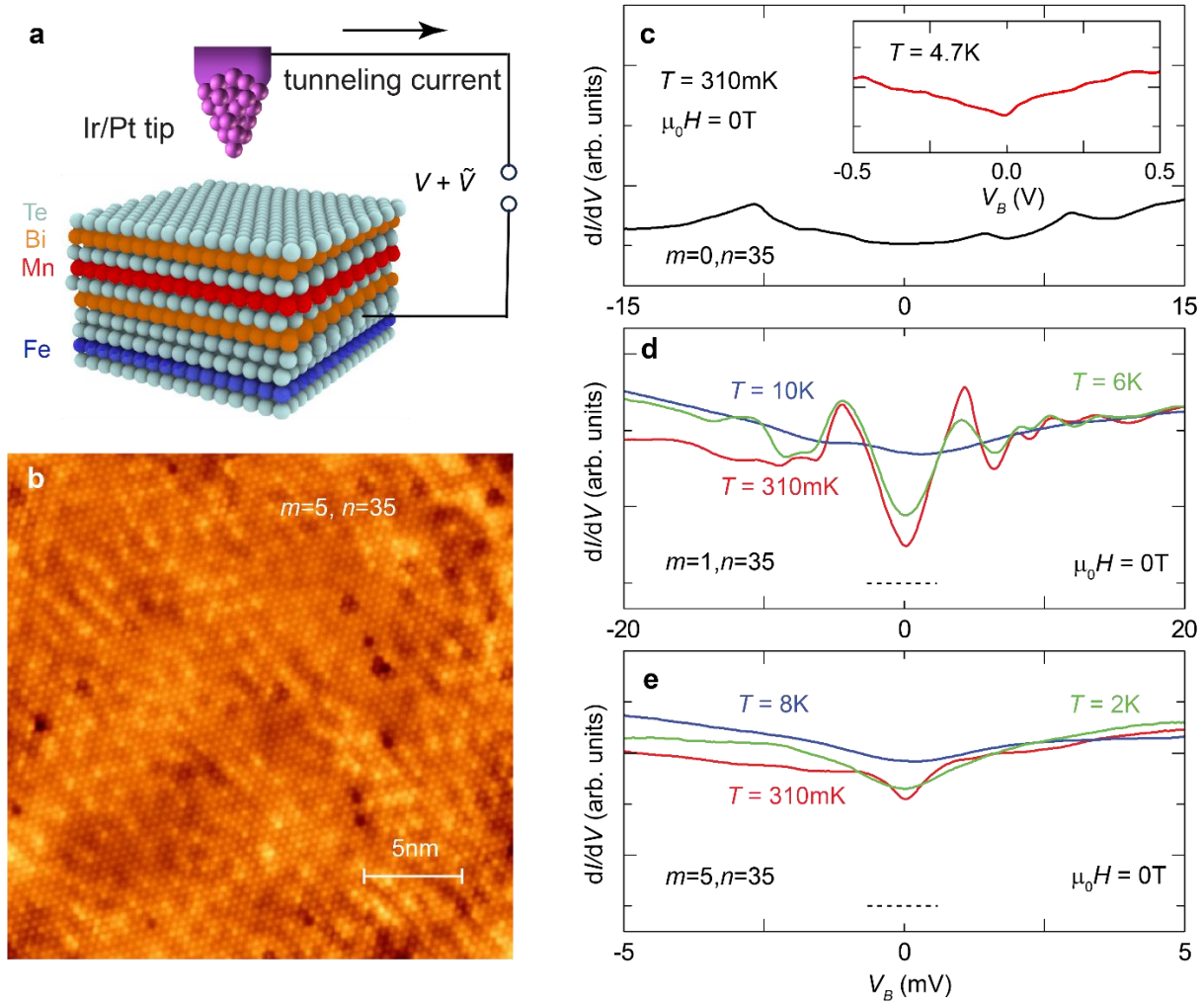

**Fig. 3| STM/S characterization of *m* SL MnBi$_2$Te$_4$/35 UC FeTe heterostructures. a,** Schematic of the STM/S measurement. **b,** Atomic resolution STM image of the (5, 35) heterostructure (sample bias $V_B$ = -300 mV, tunneling current $I_t$ = 1000 pA, $T$ = 310 mK). **c,** d$I$/d$V$ spectra on the surface of the pristine FeTe film with the range of 15 mV (setpoint: $V_B$ = +15mV, $I_t$ = 1000 pA, $T$ = 310 mK). Inset: d$I$/d$V$ spectra with the range of 0.5 V (setpoint: $V_B$ = +500 mV, $I_t$ = 500 pA, $T$ = 4.7 K). **d,** d$I$/d$V$ spectra on the top surface of the (1, 35) heterostructure at different temperatures (setpoint: $V_B$ = +20 mV, $I_t$ = 300 pA). **e,** d$I$/d$V$ spectra on the top surface of the (5, 35) heterostructure measured at different temperatures (setpoint: $V_B$ = +5 mV, $I_t$ = 300pA). The dashed lines in (**d**) and (**e**) correspond to the zero d$I$/d$V$ values of (1, 35) and (5, 35) heterostructures, respectively.

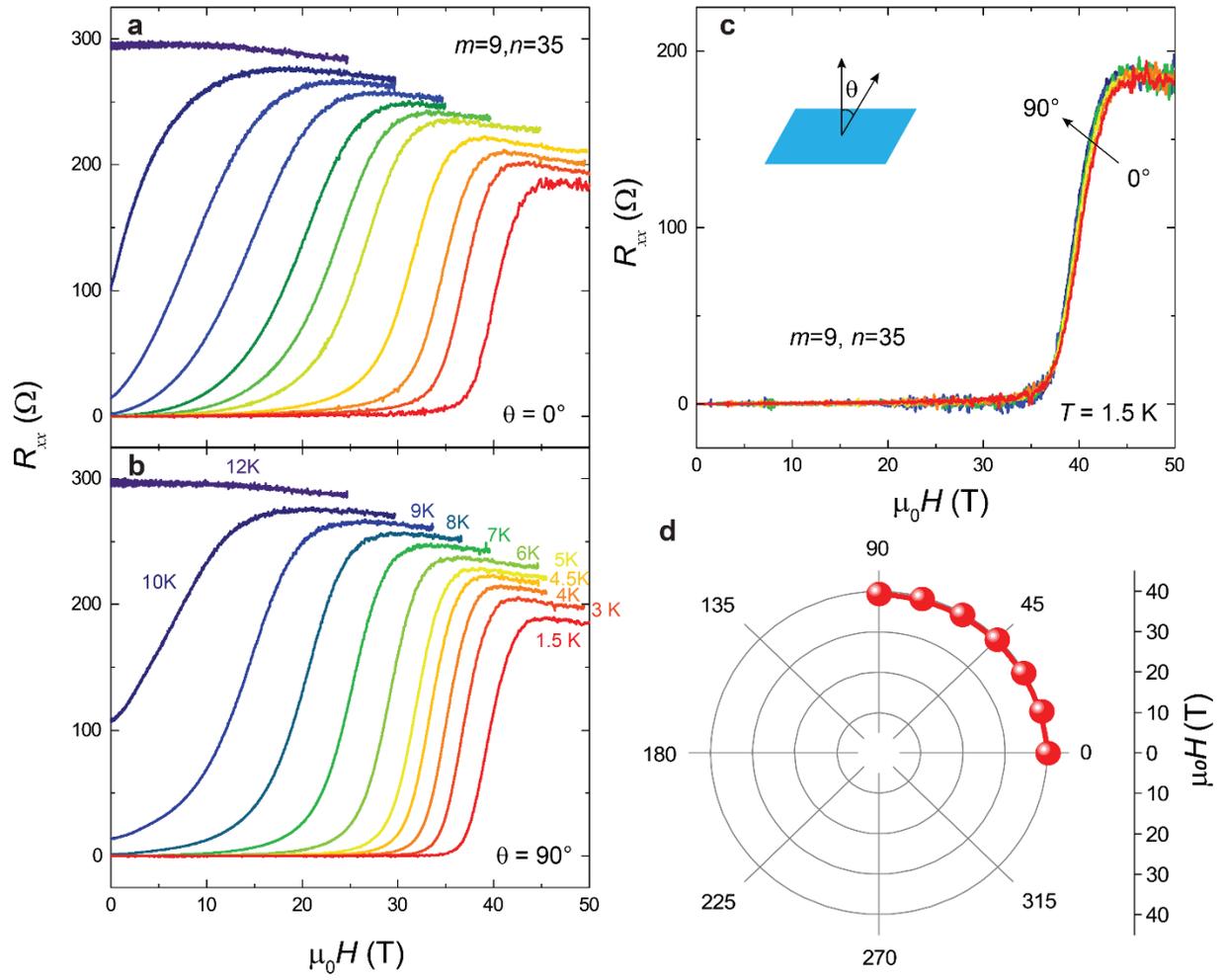

**Fig. 4| Large and isotropic upper critical magnetic field in MnBi$_2$Te$_4$/FeTe heterostructures. a, b,** $R_{xx}$-$\mu_0H$ curves of the (9, 35) heterostructure under different $T$ (**a**) $\theta = 0°$ (i.e., out-of-plane) and (**b**) $\theta = 90°$ (i.e., in-plane). **c,** $R_{xx}$-$\mu_0H$ curves of the same (9, 35) heterostructure at $T = 1.5$ K under different $\theta$. Inset shows the angle $\theta$ between the normal direction of the film and the magnetic field $\mu_0H$ direction. **d,** $\theta$ dependence of the upper critical magnetic field $\mu_0H_{c2}$. The value of $\mu_0H_{c2}$ is determined as the magnetic field at which $R_{xx}$ drops to ~50% of its normal state value.